\documentstyle[fleqn,11pt]{article}
\title{
Ouantum integrable system with two color components in two
dimensions
\thanks{The work was supported in part by National Fund of China though
C N Yang and the Grant LWTZ-1298 of Chinese Academy of Science.}}
\author{
{Mu-Lin Yan\thanks{E-mail address: mlyan@ustc.edu.cn}}
\\
Center for Fundamental Physics \\
University of Science and Technology of China\\Hefei Anhui 230026, P.R.China\\}
\date{Dec. 7, 2000}

\begin{document}
\baselineskip0.3in \maketitle
\begin{abstract}
{The Davey-Stewartson 1(DS1) system[9] is an integrable model in two
dimensions. A quantum DS1 system with 2 colour-components in two dimensions
has been formulated. This two-dimensional problem has been reduced to
two one-dimensional many-body problems with 2 colour-components.
The solutions of the two-dimensional problem under consideration has been
constructed from the resulting problems in one dimensions.
For latters with the $\delta $-function interactions and being solved by the
Bethe ansatz, we introduce symmetrical
and antisymmetrical Young operators of the permutation group  and obtain
the exact solutions for the quantum DS1 system. The application of the
solusions is discussed.}
\end{abstract}
\vskip0.05cm


\begin{center} (This paper has been published in Phys. Rev., {\bf
E61}, 4745 (2000)) \end{center}
%\begin{center} {$<$submitted to Physical Review D$>$} \end{center}
\newpage
\paragraph{1, Introduction}{\bf :}

  The Davey-Stewartson 1 (DS1) system is an integrable model in
space of two spatial and one temporal dimensions ((2+1)D). The
quantized DS1 system with scalar fields (1 component or shortly 1C)
can be formulated in  terms of the Hamiltonian of quantum
many-body problem in two dimensions, and some of them can be solved
exactly\cite{pang}\cite{yan}.
Particularly, it has been shown in ref.\cite{yan}
that these 2D quantum $N$-body system with 1C-fields can
be reduced to the solvable one-dimensional (1D) quantum $N$-body systems with
1C-fields and
with two-body potentials\cite{ols}. Thus through solving 1D quantum $N$-body
problems with 1C-fields
we can get the solutions for 2D quantum $N$-body problems with 1C-fields.
Here the key step is to separate the spatial
variables of 2D quantum $N$-body problems with 1C-fields
by constructing an ansatz\cite{pang}\cite{yan}
%\begin{equation}
$$
\Psi (\xi_1, \cdots , \xi_N,  \eta_1, \cdots , \eta_N)
= \prod_{i<j} (1-\frac{c}{4}\epsilon (\xi_{ij})\epsilon (\eta_{ij}))
  X(\xi_1, \cdots , \xi_N)Y(\eta_1, \cdots, \eta_N)
$$
%\end{equation}
where $\xi_{ij}=\xi_i-\xi_j$ and $\eta_{ij}=\eta_i-\eta_j$. This ansatz will
be called the N-body variable-separation ansatz. It is well known that
the variable-separation methods are widely used in solving high-dimensional
one-particle  problems.  For instance, for geting the wave functions of
electron in hydrgen atom, the ansatz $\Psi (r,\theta , \phi )=R(r)P(\theta )
\Phi (\phi )$ is used  to reduce the 3D problem to 1D's (this ansatz is what
we call the 1-body variable-separation ansatz).
The N-body variable-separation ansatz can be thought of as the
extension of 1-body variable-separation ansatz.
Since the N-body problems are much more complicated than
the 1-body problems, it will be highlly nontrivial to construct a
N-body variable-separation ansatz.
Ref.\cite{pang} provided the first  example for it and showed
that the idea of variable-
separating works indeed for the N-body problems  induced from
the DS1  system . \par
In this paper, we intend to generalize the above idea to multicomponents DS1
system, namely to construct a new
N-body variable-separation ansatz for the  multicomponents case  and
to solve a specific model  of 2D quantum DS1 system with  multicomponents. \par
1D N-body model with  2-components has been investigated for long
time\cite{yang}\cite{cal}. The most famous one is the model with delta-function interaction
between 2C-fermions\cite{yang}. It was  solved by the Bethe
ansatz\cite{bethe} and leads
to the Yang-Baxter equation and its thermodynamics
studies\cite{lai}\cite{sch} because of the completeness of the Bethe ansatz
solutions. In  this paper,
for definiteness, we shall study specific 2D quantum $N$-body system with
2C-fields associated with the DS1 system.
This quantum $N$-body problem under consideration can be reduced to
two 1D quantum $N$-body problems with 2C-fields
of ref.\cite{yang} and then be exactly solved by using an appropriate
N-body variable-separation ansatz and
the Bethe ansatz. \par

\paragraph{2, Quantum DS1 System with Two Components in 2-Dimension}{\bf :}

Following usual DS1 equation\cite{pang}\cite{ds}, the equation for the DS1
system with two components reads
\begin{equation}
i{\bf \dot{q}} =-\frac{1}{2} (\partial _{x}^{2}
+\partial_{y}^{2}){\bf q}+
 iA_{1}{\bf q}+iA_{2}{\bf q},
\end{equation}
where ${\bf q}$ has two colour components,
\begin{equation}
{\bf q}=
    \left( \begin{array}{c}
    q_{1}\\
    q_{2}\\
    \end{array} \right),
\end{equation}
and
\begin{eqnarray*}
(\partial _{x}-\partial
_{y})A_{1}&=&-ic(\partial_{x}+\partial_{y})({\bf q^{\dag}q})\\
(\partial _{x}+\partial
_{y})A_{2}&=&ic(\partial_{x}-\partial_{y})({\bf q^{\dag}q})
\end{eqnarray*}
where notation ${\bf \dag }$ means the hermitian transposition,
and $c$ is the coupling constant.
Introducing the coordinates $\xi =x+y, \eta =x-y,$ we have
\begin{eqnarray}
A_{1}&=&-ic\partial _{\xi}\partial _{\eta}^{-1}({\bf q^{\dag}q})
        -iu_{1}(\xi)  \\
A_{2}&=&ic\partial _{\eta}\partial _{\xi}^{-1}({\bf q^{\dag}q})
        +iu_{2}(\eta)
\end{eqnarray}
where
\begin{equation}
\partial_{\eta}^{-1}({\bf
q^{\dag}q})=\frac{1}{2}(\int_{-\infty}^{\eta} d\eta^{\prime}-
\int^{\infty}_{\eta} d\eta^{\prime}){\bf q^{\dag}}(\xi,\eta^{\prime},t)
{\bf q}(\xi,\eta^{\prime},t),
\end{equation}
and $u_{1}$ and $u_{2}$ are constants of integration. According to
ref.\cite{yan}, we choose them as
\begin{eqnarray}
u_{1}(\xi)&=&\frac{1}{2}\int d\xi^{\prime}d\eta^{\prime}U_{1}(\xi-\xi^{\prime})
           {\bf q^{\dag}}(\xi^{\prime},\eta^{\prime},t)
               {\bf q}(\xi^{\prime},\eta^{\prime},t)  \\
u_{2}(\eta)&=&\frac{1}{2}\int d\xi^{\prime}d\eta^{\prime}U_{2}(\eta-\eta^{\prime})
           {\bf q^{\dag}}(\xi^{\prime},\eta^{\prime},t)
               {\bf q}(\xi^{\prime},\eta^{\prime},t).
\end{eqnarray}
Thus eq.(1) can be written as
\begin{eqnarray}
i{\bf \dot{q}}=-(\partial_{\xi}^{2}+\partial_{\eta}^{2})
  {\bf q}+c[\partial_{\xi}\partial_{\eta}^{-1}({\bf q^{\dag}q})
+\partial_{\eta}\partial_{\xi}^{-1}({\bf q^{\dag}q})]{\bf q}
\nonumber \\
+\frac{1}{2}\int d\xi^{\prime}d\eta^{\prime}[U_{1}(\xi-\xi^{\prime})+
+U_{2}(\eta-\eta^{\prime})]({\bf q}^{\dag \prime}{\bf
q}^{\prime}){\bf q},
\end{eqnarray}
where ${\bf q}^{\prime}={\bf q}(\xi^{\prime},\eta^{\prime},t)$. We
quantize the system with the canonical commutation relations
%\begin{eqnarray}
\begin{equation}
 [ q_{a}(\xi,\eta,t), q^{\dag}_{b}(\xi^{\prime},
\eta^{\prime},t)]_{\pm}
= 2\delta_{_{ab}}\delta(\xi-\xi^{\prime})\delta(\eta-\eta^{\prime}), \\
\end{equation}
\begin{equation}
 [ q_{a}(\xi,\eta,t), q_{b}(\xi^{\prime},\eta^{\prime},t)]_{\pm}
=0.
\end{equation}
%\end{eqnarray}
where $a,b = 1$ or $2$, $[ , ]_{+}$ and $[ , ]_{-}$ are anticommutator
 and commutator respectively.
Then eq.(8) can be written in the form
\begin{equation}
\dot{{\bf q}}=i[H,{\bf q}]
\end{equation}
where $H$ is the Hamiltonian of the system
\begin{eqnarray}
H=\frac{1}{2}\int d\xi d\eta \begin{array}{c}${\LARGE(}$\end{array}
 -{\bf q}^{\dag}
(\partial_{\xi}^{2}+\partial_{\eta}^{2})
  {\bf q}+\frac{c}{2}{\bf q}^{\dag}[(\partial_{\xi}\partial_{\eta}^{-1}
  +\partial_{\eta}\partial_{\xi}^{-1})({\bf q^{\dag}q})]{\bf q}
\nonumber \\
+\frac{1}{4}\int d\xi^{\prime}d\eta^{\prime}{\bf q}^{\dag}
[U_{1}(\xi-\xi^{\prime})
+U_{2}(\eta-\eta^{\prime})]({\bf q}^{\prime \dag}{\bf
q}^{\prime}){\bf q}\begin{array}{c}${\LARGE)}$\end{array}.
\end{eqnarray}
The $N$-particle eigenvalue problem is
\begin{equation}
H\mid \Psi \rangle =E\mid \Psi \rangle
\end{equation}
where
%\begin{eqnarray}
\begin{equation}
\mid \Psi \rangle=\int d\xi_{1}d\eta_{1}\ldots d\xi_{N}d\eta_{N}
\sum_{a_{1}\ldots a_{N}} \Psi_{a_{1}\ldots a_{N}}(\xi_{1}\eta_{1}
\ldots \xi_{N}\eta_{N})
% \nonumber  \\
{\bf q}^{\dag}_{a_{1}}(\xi_{1}\eta_{1})\ldots
{\bf q}^{\dag}_{a_{N}}(\xi_{N}\eta_{N}) \mid 0 \rangle.
%\end{eqnarray}
\end{equation}
The $N$-particle wave function $ \Psi_{a_{1}\ldots a_{N}}$ is defined
by eq.(14), which satisfies the $N$-body Schr\"{o}dinger equation
\begin{eqnarray}
-\sum_{i}(\partial_{\xi_{i}}^{2}+\partial_{\eta_{i}}^{2})
 \Psi_{a_{1}\ldots a_{N}}+c\sum_{i<j}
 [\epsilon (\xi_{ij})\delta^{\prime}(\eta_{ij})
  +\epsilon (\eta_{ij})\delta^{\prime}(\xi_{ij})] \Psi_{a_{1}\ldots a_{N}}
  \nonumber  \\
  +\sum_{i<j}[U_{1}(\xi_{ij})+U_{2}(\eta_{ij})] \Psi_{a_{1}\ldots a_{N}}
  =E \Psi_{a_{1}\ldots a_{N}}
\end{eqnarray}
where $\xi_{ij}=\xi_{i}-\xi_{j}, \delta^{\prime}(\xi_{ij})=\partial_{\xi_{i}}
\delta (\xi_{ij})$, and $\epsilon (\xi_{ij})=1$ for $\xi_{ij}>0, 0$ for
$\xi_{ij}=0, -1$ for $\xi_{ij}<0.$ Since there are products of
distributions in eq.(15), an appropriate regularezation for avoiding
uncertainty is necessary. This issue has been discussed in ref.\cite{zhao}.

\paragraph{3, Variable Separation of Quantum DS1 with Two Components
and Bethe Ansatz}{\bf :}

  Our purpose is to solve the $N$-body Schr\"{o}dinger equation (15). The
results in ref.\cite{yan} remind us that we can make the following ansatz
\begin{eqnarray}
\Psi_{a_{1}\ldots a_{N}}
&=&\sum_{_{\begin{array}{ccc}
       a_{1}^{\prime} & \ldots & a_{N}^{\prime} \\
       b_{1}^{\prime} & \ldots & b_{N}^{\prime} \end{array}}}
 \prod_{i<j} (1-\frac{c}{4}\epsilon (\xi_{ij})\epsilon (\eta_{ij}))
 \nonumber  \\
& &\times {\cal M}_{a_{1}\ldots a_{N}, a_{1}^{\prime}\ldots a_{N}^{\prime}}
 {\cal N}_{a_{1}\ldots a_{N}, b_{1}^{\prime}\ldots b_{N}^{\prime}}
X_{ a_{1}^{\prime}\ldots a_{N}^{\prime}}(\xi_{1}\ldots \xi_{N})
Y_{ b_{1}^{\prime}\ldots b_{N}^{\prime}}(\eta_{1}\ldots \eta_{N})
\end{eqnarray}
where ${\cal M}$ and ${\cal N}$ are matrices being independent of
$\xi$ and $\eta$,
and both $X_{ a_{1}\ldots a_{N}}(\xi_{1}\ldots \xi_{N})$ and
$Y_{ b_{1}\ldots b_{N}}(\eta_{1}\ldots \eta_{N})$ are one-dimensional wave
functions of N-bodies. Substituting eq.(16) into eq.(15), we abtian
\begin{eqnarray}
-\sum_{i} \partial_{\xi_{i}}^{2} X_{a_{1}\ldots a_{N}}
+\sum_{i<j} U_{1}(\xi_{ij}) X_{a_{1}\ldots a_{N}}&=&E_{1} X_{a_{1}\ldots a_{N}}
\\
-\sum_{i} \partial_{\eta_{i}}^{2} Y_{b_{1}\ldots b_{N}}
+\sum_{i<j} U_{2}(\eta_{ij}) Y_{b_{1}\ldots b_{N}}&=&E_{2}Y_{b_{1}\ldots b_{N}}
\end{eqnarray}
where $U_{1}(\xi_{ij})$ and $U_{2}(\eta_{ij})$ are two-body potentials, eqs.
(17) (18) are one-dimensional $N$-body Schr\"{o}dinger equations and
$E_{1}+E_{2}=E$. Above derivation indicates that the two-dimensional $N$-body
Schr\"{o}dinger equation (15) has been reduced into two one-dimentional
 $N$-body Schr\"{o}dinger equations. Namely, the variables in the
two-dimentional $N$-body wave function $\Psi_{a_{1}\ldots a_{N}}$ have been
separated.

At this stage ${\cal M}$ and ${\cal N}$ are
unknown temporarily.
 It is expected that for any given pair of exactly solvable 1D N-body
 problems and the  correspondent solutions,
we could construct the solutions $\Psi
_{A_{1} \ldots A_{N}}$ for 2D N-body problems eq.(15) through
constructing an appropriate ${\cal M} \times {\cal N}$-matrix.
% where $\otimes$ means the tenser product.
It has been known that the 1D  N-body problem in the form of (17) or (18)
can be solved exactly for a class of
potentials\cite{yang}\cite{cal}\cite{sut}.  To illustrate
the  construction of ${\cal M} \times {\cal N}$-matrix, we  take  both potentials
in  (17) and (18) the delta functions
$U_{1}(\xi_{ij})=2g\delta (\xi_{ij})$
and $U_{2}(\eta_{ij})=2g\delta (\eta_{ij})$ ($g>0$, the coupling
constant). Then eqs.(17) and (18) become
\begin{eqnarray}
-\sum_{i} \partial_{\xi_{i}}^{2} X_{a_{1}\ldots a_{N}}
+2g \sum_{i<j} \delta(\xi_{ij}) X_{a_{1}\ldots a_{N}}
&=&E_{1} X_{a_{1}\ldots a_{N}}
\\
-\sum_{i} \partial_{\eta_{i}}^{2} Y_{b_{1}\ldots b_{N}}
+2g \sum_{i<j} \delta(\eta_{ij}) Y_{b_{1}\ldots b_{N}}
&=&E_{2}Y_{b_{1}\ldots b_{N}}.
\end{eqnarray}
As $X$ and $Y$ are wave functions of Fermions with two components,
denoted by $X^{F}$ and $Y^{F}$, the
problem has been solved by Yang long ago\cite{yang} (more explicitly, see
ref.\cite{gu} and ref.\cite{fan}). According to the Bethe ansatz, the continual solution of
eq.(9) in the region of $0<\xi_{Q_{1}}<\xi_{Q_{2}}<\ldots
<\xi_{Q_{N}}<L$ reads
\begin{eqnarray}
X^{F}&=&\sum_{P}
\alpha_{P}^{(Q)}\exp\{i[k_{P_{1}}\xi_{Q_{1}}+\ldots+k_{P_{N}}\xi_{Q_{N}}
]\} \nonumber  \\
&=&\alpha_{_{12\ldots N}}^{(Q)}e^{i(k_{1}\xi_{Q_{1}}+k_{2}\xi_{Q_{2}}
+\ldots+k_{N}\xi_{Q_{N}})}
+\alpha_{_{21\ldots N}}^{(Q)}e^{i(k_{2}\xi_{Q_{1}}+k_{1}\xi_{Q_{2}}
+\ldots+k_{N}\xi_{Q_{N}})}
\nonumber  \\
& &+(N!-2) \begin{array}{c} $others terms$ \end{array}
\end{eqnarray}
where  $X^{F}\in \{ X^{F}_{a_{1}\ldots a_{N}} \}$, $P=[P_{1},P_{2},\cdots,P_{N}]$
and $Q=[Q_{1},Q_{2},\cdots,Q_{N}]$
are two permutations of the
integers $1,2,\ldots ,N$, and
\begin{eqnarray}
\alpha^{(Q)}_{\ldots ij \ldots}&=&Y^{lm}_{ji} \alpha^{(Q)}_{\ldots ji \ldots},
  \\
Y^{lm}_{ji}&=&\frac{-i(k_{j}-k_{i})P^{lm}+g}{i(k_{j}-k_{i})-g}
\end{eqnarray}
The eigenvalue is given by
\begin{equation}
E_{1}=k_{1}^{2}+k_{2}^{2}+\ldots +k_{N}^{2},
\end{equation}
where $\{ k_{i} \}$ are determined by the Bethe ansatz equations,
\begin{eqnarray}
& &e^{ik_{j}L}=\prod_{\beta =1}^{M} \frac{i(k_{j}-\Lambda_{\beta})-g/2}
{i(k_{j}-\Lambda_{\beta})+g/2}   \\
& &\prod_{j=1}^{N} \frac{i(k_{j}-\Lambda_{\alpha})-g/2}
{i(k_{j}-\Lambda_{\alpha})+g/2}
=-\prod_{\beta =1}^{M} \frac{i(\Lambda_{\alpha}-\Lambda_{\beta})+g}
{i(\Lambda_{\alpha}-\Lambda_{\beta})-g}
\end{eqnarray}
with $\alpha =1,\ldots, M, j=1,\ldots, N.$ Through exactly same
procedures we can get the solution $Y^{F}$ and $E_{2}$ to eq.(20).

    As $X$ and $Y$ are Boson's wave-functions,
denoted by $X^{B}$ and $Y^{B}$, it
is easy to be shown that
\begin{eqnarray}
X^{B}&=&\sum_{P}
\beta_{P}^{(Q)}\exp\{i[k_{P_{1}}\xi_{Q_{1}}+\ldots+k_{P_{N}}\xi_{Q_{N}}
]\}   \\
\beta^{(Q)}_{\ldots ij \ldots}&=&Z^{lm}_{ji} \beta^{(Q)}_{\ldots ji \ldots},
  \\
Z^{lm}_{ji}&=&\frac{i(k_{j}-k_{i})P^{lm}+g}{i(k_{j}-k_{i})-g}
\end{eqnarray}
and the Bethe ansatz equations are as following\cite{fan}
\begin{eqnarray}
& &e^{ik_{j}L}=(-1)^{N+1}\prod_{i=1}^{N} \frac{k_{j}-k_{i}+ig}{k_{j}-k_{i}-ig}
\prod_{\beta =1}^{M} \frac{\Lambda_{\beta}-k_{j}+ig/2}
{\Lambda_{\beta}-k_{j}-ig/2}   \\
& &\prod_{\alpha =1}^{M} \frac{\Lambda_{\beta}-\Lambda_{\alpha}+ig}
{\Lambda_{\beta}-\Lambda_{\alpha}-ig}
=(-1)^{M+1}\prod_{j =1}^{N} \frac{\Lambda_{\beta}-k_{j}+ig/2}
{\Lambda_{\beta}-k_{j}-ig/2}.
\end{eqnarray}
$Y^{B}$ is same as $X^{B}$. It is well known that $X^{F}$ and
$Y^{F}$($X^{B}$ and $Y^{B}$) are antisymmetrical (symmetrecal) as
the coordinates and the colour-indeces of the
particles interchanges each other simultaneously, instead of
the coordinates interchanges each other merely.

\paragraph{4, Young Operator of Permutation Group}{\bf :}

     For permutation group $S_{N}: \{ e_{i}, i=1,\cdots,
N!\}$, the totally symmetrical Young operator is
\begin{equation}
{\cal O}_{N}=\sum_{i=1}^{N!} e_{i},
\end{equation}
and the totally antisymmetrical Young operator is
\begin{equation}
{\cal A}_{N}=\sum_{i=1}^{N!} (-1)^{P_{i}}e_{i}.
\end{equation}
The Young diagram for ${\cal O}_{N}$ is
\begin{tabular}{|l|l|l|l|l|r|}  \hline
 1  & 2 &3 &$\cdots$ & N  \\  \hline
\end{tabular},
and for ${\cal A}_{N}$, it is
\begin{tabular}{|l|r|}  \hline
 1  \\  \hline
 2  \\  \hline
 \vdots  \\  \hline
 N  \\   \hline
\end{tabular}.
To $S_{3}$, for example, we have
\begin{equation}
{\cal O}_{3}=1+P^{12}+P^{13}+P^{23}+P^{12}P^{23}+P^{23}P^{12}.
\end{equation}
\begin{equation}
{\cal A}_{3}=1-P^{12}-P^{13}-P^{23}+P^{12}P^{23}+P^{23}P^{12}.
\end{equation}
Lemma 1: $({\cal O}_{N}X_{F})(\xi_{1},\xi_{2},\cdots,\xi_{N})$ is
antisymmetrical with respect to the coordinate's interchanges of
$(\xi_{i} \longleftrightarrow \xi_{j})$.

\noindent Proof: From the definition of ${\cal O}_{N}$ (eq.(32)), we
have
\begin{equation}
{\cal O}_{N}P^{ab}=P^{ab}{\cal O}_{N}={\cal O}_{N}.
\end{equation}
To $N=3$ case, for example, the direct calculations show ${\cal
O}_{3}P^{12}=P^{12}{\cal O}={\cal O}_{3}, {\cal
O}_{3}P^{23}=P^{23}{\cal O}={\cal O}_{3}$
and so on. Using eqs.(36) and (23), we have
\begin{equation}
{\cal O}_{N}Y^{lm}_{ij}=(-1){\cal O}_{N}.
\end{equation}
From eqs.(21) and (23), $X^{F}$ can be written as
\begin{eqnarray}
X^{F}&=&\{ e^{i(k_{1}\xi_{Q_{1}}+k_{2}\xi_{Q_{2}}
+\ldots+k_{N}\xi_{Q_{N}})}
+Y^{12}_{12}e^{i(k_{2}\xi_{Q_{1}}+k_{1}\xi_{Q_{2}}
+\ldots+k_{N}\xi_{Q_{N}})}
\nonumber  \\
& &+Y^{23}_{13}Y^{12}_{12}e^{i(k_{2}\xi_{Q_{1}}
+k_{3}\xi_{Q_{2}}+k_{1}\xi_{Q_{3}}
+\ldots+k_{N}\xi_{Q_{N}})}
+(N!-3) \begin{array}{c} $other terms$ \end{array}
\} \alpha_{_{12\ldots N}}^{(Q)}
\end{eqnarray}
Using eqs.(37) and (38), we obtain
\begin{eqnarray}
({\cal O}_{N}X^{F})(\xi_{1},\cdots,\xi_{N})=
\{ e^{i(k_{1}\xi_{Q_{1}}+k_{2}\xi_{Q_{2}}
+\ldots+k_{N}\xi_{Q_{N}})}
-e^{i(k_{2}\xi_{Q_{1}}+k_{1}\xi_{Q_{2}}
+\ldots+k_{N}\xi_{Q_{N}})}
\nonumber   \\
+e^{i(k_{2}\xi_{Q_{1}}
+k_{3}\xi_{Q_{2}}+k_{1}\xi_{Q_{3}}
+\ldots+k_{N}\xi_{Q_{N}})}
+(N!-3)\begin{array}{c}$other terms$\end{array}
\} {\cal O}_{N} \alpha_{_{12\ldots N}}^{(Q)}
\nonumber   \\
=\sum_{P} (-1)^{P}
\exp\{i[k_{P_{1}}\xi_{Q_{1}}+\ldots+k_{P_{N}}\xi_{Q_{N}}
]\}({\cal O}_{N} \alpha_{_{12\ldots N}}^{(Q)}).
\end{eqnarray}
Therefore we conclude that $({\cal O}_{N}X^{F})(\xi_{1},\cdots,\xi_{N})$
is antisymmetrical
with respect to $(\xi_{i}\longleftrightarrow \xi_{j})$.

\noindent Lemma 2: $({\cal A}_{N}X^{B})(\xi_{1},\xi_{2},\cdots,\xi_{N})$ is
antisymmetrical with respect to the coordinate's interchanges of
$(\xi_{i} \longleftrightarrow \xi_{j})$.

\noindent  Proof: Noting (see eqs.(33) (29) (27))
\begin{eqnarray}
{\cal A}_{N} P^{ab}=P^{ab}{\cal A}=-{\cal A}_{N},  \\
{\cal A}_{N} Z^{lm}_{ij}=(-1){\cal A}_{N},
\end{eqnarray}
we then have
\begin{equation}
({\cal A}_{N}X^{B})(\xi_{1},\cdots,\xi_{N})
=\sum_{P} (-1)^{P}
\exp\{i[k_{P_{1}}\xi_{Q_{1}}+\ldots+k_{P_{N}}\xi_{Q_{N}}
]\}({\cal A}_{N} \beta_{_{12\ldots N}}^{(Q)}).
\end{equation}
Then the Lemma is proved.

\paragraph{5, The Solutions of the Problem}{\bf :}

    The ansatz of eq.(16) can be compactly written as
\begin{equation}
\Psi
= \prod_{i<j} (1-\frac{c}{4}\epsilon (\xi_{ij})\epsilon (\eta_{ij}))
  ({\cal M}X)  ({\cal N}Y)
\end{equation}
where $({\cal M}X)$ and $({\cal N}Y)$ are required to be
antisymmetrical under the interchanges of the coodinate vairables.
According to Lemmas 1 and 2, we see that
\begin{equation}
 {\cal M},{\cal N} =\left\{
\begin{array}{ccc}{\cal O}_{N} & &$for 1D Fermion$  \\
                  {\cal A}_{N} & &$for 1D Boson.$
\end{array} \right.
\end{equation}
As the DS1 fields $q_{a}(\xi \eta)$ in eq.(1) are (2+1)D Bose fields,
the commutators ($[ , ]_{-}$, see (9) and (10)) are used to quantized
the system and the 2D N-body wave functions denoted in $\Psi ^{B}$
must be symmetrical under the colour-interchang $(a_{i} \longleftrightarrow
a_{j})$ and the coordinate-interchange $((\xi_{i} \eta_{i})
\longleftrightarrow  (\xi_{j} \eta_{i}))$. Namly, the 2D Bose wave
functions $\Psi^{B}$ must satisfy that
\begin{equation}
P^{a_{i} a_{j}}\Psi^{B}\mid _{\xi_{i}\eta_{i} \longleftrightarrow
\xi_{j} \eta_{j}} =\Psi^{B}.
\end{equation}
As $q_{a}$ are (2+1)D Fermi fields, the anticommutators should be used,
and $\Psi^{F}$ must be antisymmetrical under $(a_{i} \longleftrightarrow
a_{j})$ and $((\xi_{i} \eta_{i})
\longleftrightarrow  (\xi_{j} \eta_{i}))$. Namly,
\begin{equation}
P^{a_{i} a_{j}}\Psi^{F}\mid _{\xi_{i}\eta_{i} \longleftrightarrow
\xi_{j} \eta_{j}} =-\Psi^{F}.
\end{equation}
Thus for the 2D Boson case, two solutions of $\Psi^{B}$ can be
constructed as following
\begin{eqnarray}
\Psi^{B}_{1}
= \prod_{i<j} (1-\frac{c}{4}\epsilon (\xi_{ij})\epsilon (\eta_{ij}))
 [ {\cal O_{N}}X^{F}(\xi_{1}\cdots \xi_{N})]
 [ {\cal O_{N}}Y^{F}(\eta_{1}\cdots \eta_{N})], \\
\Psi^{B}_{2}
= \prod_{i<j} (1-\frac{c}{4}\epsilon (\xi_{ij})\epsilon (\eta_{ij}))
 [ {\cal A_{N}}X^{B}(\xi_{1}\cdots \xi_{N})]
  [{\cal A_{N}}Y^{B}(\eta_{1}\cdots \eta_{N})].
\end{eqnarray}
Using eqs.(36),(39),(40) and (42), we can check eq.(45) diractly. In
addition, from the Bethe ansatz equations (25) (26) (30) (31) and $E=
E_{1}+E_{2}$, we can see that the eigenvalues of $\Psi^{B}_{1}$ and
$\Psi^{B}_{2}$ are different each other generally, i.e., the states
corrosponding to $\Psi^{B}_{1}$ and $\Psi^{B}_{2}$ are non-degenerate.

For the 2D Fermion case, the desired results are
\begin{eqnarray}
\Psi^{F}_{1}
= \prod_{i<j} (1-\frac{c}{4}\epsilon (\xi_{ij})\epsilon (\eta_{ij}))
 [ {\cal O_{N}}X^{F}(\xi_{1}\cdots \xi_{N})]
 [ {\cal A_{N}}Y^{B}(\eta_{1}\cdots \eta_{N})], \\
\Psi^{F}_{2}
= \prod_{i<j} (1-\frac{c}{4}\epsilon (\xi_{ij})\epsilon (\eta_{ij}))
 [ {\cal A_{N}}X^{B}(\xi_{1}\cdots \xi_{N})]
  [{\cal O_{N}}Y^{F}(\eta_{1}\cdots \eta_{N})].
\end{eqnarray}
Eq.(46) can also be checked diractly. The eigenvalues corresponding to
$\Psi^{F}$ are also determined by the Bethe equations and
$E=E_{1}+E_{2}$.

 It is similar to ref.\cite{yan} that we can prove $\Psi^{B}_{1,2}$ and
$\Psi^{F}_{1,2}$ shown in above are of
the exact solutions of the eq.(15). Thus
 we conclude that the 2D quantum
many-body problem induced from the quantum DS1 system with
2-component has been solved exactly.

\paragraph{6, The Ground-State Energies of the System}{\bf :}

In this section, we discuss the ground-state energies of the DS1 system
solved in the previous section by using the Bethe ansatz equations
(25), (26) and (30), (31). Let the 2D N-body problem reduced from 2D DS1
system with 2 colour (or spin) components has $M$ colours down and
$N-M$ colours up. Therefore both $X^{F,B}(\xi_{1},\xi_{2},\cdots \xi_{N})$ and
$Y^{F,B}(\eta_{1}, \eta_{2},\cdots \eta_{N})$ in eqs(47)-(50) are one
dimensional $N-$body wave functions with $M$ colours down and $N-M$
colours up.
We are interested in the limit that $N$, $M$ and the length $L$ of the box
go to infinity proportionately, i.e., both $N/L=D$ and $M/L=D_m$ are finite.

For one dimensional $N$-fermion problem, by the nested Bethe ansatz (or
Bethe-Yang ansatz) equations (25) and (26), the corresponding integration
equations for the ground state read\cite{yang}
\begin{eqnarray}
2\pi\sigma_1&=&-\int_{-B_1}^{B_1}{{2g\sigma_1(\Lambda')d\Lambda'}\over
{g^2+(\Lambda-\Lambda')^2}}
+\int_{-Q_1}^{Q_1}{{4g\rho_1(k)dk}\over
{g^2+4(k-\Lambda)^2}},\\
2\pi\rho_1&=&1
+\int_{-B_1}^{B_1}{{4g\sigma_1(\Lambda)d\Lambda}\over
{g^2+4(k-\Lambda)^2}},
\end{eqnarray}
where $\rho_1(k)$ is particle (i.e.,1D fermion) density distribution
function of $k$, and $\sigma_1(\Lambda)$ is colour-down particle density
distribution function of $\Lambda$. Namely, we have
\begin{equation}
D=\int_{-Q_1}^{Q_1}{\rho_1(k)dk},\;\;\;
D_m=\int_{-B_1}^{B_1}{\sigma_1(\Lambda)d\Lambda},\;\;\;
E_1/N=D^{-1}\int_{-Q_1}^{Q_1}{k^2\rho_1(k)dk}.
\end{equation}

For 1D N-boson case, starting from the nested Bethe ansatz equations (30)
and (31), similar integration equations for ground state of bosons can be
derived (see Appendix). The results are as follows
\begin{eqnarray}
2\pi\sigma_2&=&\int_{-B_2}^{B_2}{{2g\sigma_2(\Lambda')d\Lambda'}\over
{g^2+(\Lambda-\Lambda')^2}}
-\int_{-Q_2}^{Q_2}{{4g\rho_2(k)dk}\over
{g^2+4(k-\Lambda)^2}},\\
2\pi\rho_2&=&1
-\int_{-B_2}^{B_2}{{4g\sigma_2(\Lambda)d\Lambda}\over
{g^2+4(\Lambda-k)^2}}+\int_{-Q_2}^{Q_2}{{2g\rho_2(k')dk'}\over
{g^2+(k-k')^2}},
\end{eqnarray}
where $\rho_2(k)$ and $\sigma_2(\Lambda)$ are bosonic particle density
distribution function of $k$ and its colour-down particle density
distribution function of $\Lambda$ respectively, i.e.,
\begin{equation}
D=\int_{-Q_2}^{Q_2}{\rho_2(k)dk},\;\;\;
D_m=\int_{-B_2}^{B_2}{\sigma_2(\Lambda)d\Lambda},\;\;\;
E_2/N=D^{-1}\int_{-Q_2}^{Q_2}{k^2\rho_2(k)dk}.
\end{equation}

The everage energies of the 2D DS1 ground states described by $\Psi^B_1$,
$\Psi^B_2$, $\Psi^F_1$ and $\Psi^F_2$ (see eqs(47)$-$(50)) are denoted
by $E(\Psi^B_1)$, $E(\Psi^B_2)$, $E(\Psi^F_1)$ and $E(\Psi^F_2)$
respectively. Then, the everage energies per particle for the ground-states
are follows
\begin{eqnarray}
E(\Psi^B_1)/N&=&2E_1/N=2D^{-1}\int_{Q_1}^{Q_1}k^2\rho_1(k)dk, \\
E(\Psi^B_2)/N&=&2E_2/N=2D^{-1}\int_{Q_2}^{Q_2}k^2\rho_2(k)dk, \\
E(\Psi^F_1)/N&=&{1\over N} (E_1+E_2) \nonumber \\
  &=&D^{-1}(\int_{Q_1}^{Q_1}k^2\rho_1(k)dk+\int_{Q_2}^{Q_2}k^2\rho_2(k)dk)
  \nonumber  \\
  &=&{1\over 2}(E(\Psi^B_1)+E(\Psi^B_2)), \\
E(\Psi^F_2)/N&=&E(\Psi^F_1)/N.
\end{eqnarray}
From these equations, the follows can been seen:
1, The everage energies per particle
for the ground states of this 2-dimensional (2D-) DS1
problem are reduced into the everage energies per particle of 1-dimensional
(1D-)many body problems.
As $D$ and $D_m$ are given, by solving the integration
equations $(51)-(56)$, we obtain the $\rho_1(k)$ and $\rho_2(k)$,
and then get the desired results of $E(\Psi^B_1)/N$, $E(\Psi^B_2)/N$,
$E(\Psi^F_1)/N$ and $E(\Psi^F_2)/N$.
2, For the two bosonic
solutions of the 2D DS1 system with 2 colours (eqs (47) (48)), the everage
ground state energies per particle are twice as large as one of 1D-fermions
or 1D-bosons;
3, For the fermion solutions of this 2D-DS1 system, $E(\Psi^F_1)/N$
and $E(\Psi^F_2)/N$ are sum of 1D-fermion everage energy per particle
and 1D-boson's.
4, In general, $E(\Psi^B_1)\neq E(\Psi^B_2)\neq E(\Psi^F_{1,\;{\rm or}\;2})$.
Namely, for same DS1 system, if the statistics of the wave functions
(or particles) is different, the corresponding ground-state energies are
different. This is remarkable and reflects the statistical effects
in the 2D DS1 system.

\paragraph{7, Discussions and Summary}{\bf :}

Finally, we would like to speculate some further applications of the results
presented in this paper to the mathematical physics. Our results may be useful
in the following two respects. Firstly, the Bethe ansatz equations (25), (26)
for fermion wave functions and (30), (31) for boson's can be solved
respectively, even though the equations are systems of transcendental
equations for which the roots are not easy to locate. The so-called string
hypothesis is used for the analysis and classification of the roots for
the Bethe ansatz equations\cite{lai}\cite{sch}. Thus, we could study thier
ground state, the excitation and the thermodynamics based on
it\cite{lai}\cite{sch}. Then, the thermodynamical properties of the 1D Bose
or Fermi gas with $\delta-$function interaction and with two components can
be explored. The eqs.(47)$-$(50) indicate that under the thermodynamical
limit the 2D DS1 gases (with two colour components) are classified into
2D Bose gases and 2D Fermi gases.
By eqs (47) (48), the 2D Bose gases are composeted of two 1D Fermi
gases or 1D Bose gases, and by eqs (49) (50), the 2D Fermi gases are
composeted of 1D Fermi gas and 1D Bose gas. Hence, the thermodynamics of
2D DS1 gases with two colour components can be derived exactly. It would
be interesting in physics, because this is an interesting and nontrivial
example to illustrate coupling (or fusing) of two 1D 2-component gases
with $\delta-$function interactin and with different or same statistics.
Secondly, the colourless DS1 equation originated in studies of nonlinear
phenomena\cite{ds}. Five years ago, Pang, Pu and Zhao\cite{ppz} showed an
example that the solutions of the initial-boundary-value problem for the
related classical DS1 equation in ref.\cite{fokas} are consistent with the
solutions for the quantum DS1 system with time-dependent applied forces.
This indicates that the classical solutions of DS1 equation are corresponding
to the classcal limit of the solutions for the quantum DS1 system.
This is actully a new method to reveal the solutions of the colourless
DS1 equation.
To the quantum DS1 system with colour indices studied in this present paper,
similar correspondences are expectable. Hence, the structure of the
solutions of the quantum DS1-system with colour indices revealed in this
paper would be helpful to understand the corresponding classical
solutions of DS1 systems with colour. The specific studies on the above
speculations would be meaningful, however, they are beyond the scope of
this present paper.

To summarize. We formulated the quantum multicomponent DS1 system in
terms of the quantum multicomponent many-body Hamiltonain in 2D space.
Then we reduced this 2D Hamiltonain to two 1D multicomponent many-body
problems. As the potential between two particles with two components in
one dimension is $\delta-$function, the Bethe ansatz was used to solve
these 1D problems. By using the ansatz of ref.\cite{pang} and
introducing some useful Young
operators, we presented a new N-body variable-separation ansatz
for fusing two 1D-solutions
to construct 2D wave functions of the quantum many-body problem which is
induced from the quantum 2-component DS1 system. There are two types of
wave functions: Boson's and Fermion's. Both of them satisfy the 2D many-body
Schr\"{o}dinger equation of the DS1 system exactly.
The results have been used to study the ground states of the system.
Some further applications of
the results presented in this paper are speculated and discussed.

\begin{center}  {\bf Acknowledgement}   \end{center}

The author would like to thank Y.Chen and B.H.Zhao for thier helpful
discussions.

\vskip0.5cm

%\newpage

\begin{center}  {\bf Appendix}   \end{center}
Let us derive eqs (54) and (55) in the text. We start from the Bethe ansatz
equations (30) and (31) of 1D bosons with two colour componts. Taking the
logarithm of (30) and (31) respectively, we have
\begin{eqnarray*}
&&k_jL=2\pi I_k-2\sum\limits_{i=1}^{N}\tan^{-1}{{k_j-k_i}\over{g}}
      -2\sum\limits_{\beta=1}^{M}\tan^{-1}{{2(\Lambda_\beta-k_j)}\over{g}}
%\begin{flushright} (A1) \end{flushright}\\
      \hspace{1.5in}(A1)\\
&&2\sum\limits_{\alpha=1}^{M}\tan^{-1}{{\Lambda_\beta-\Lambda_\alpha}
\over{g}}=2\pi J_\Lambda
+2\sum\limits_{j=1}^{N}\tan^{-1}{{2(\Lambda_\beta-k_j)}\over{g}},
\hspace{1.3in}(A2)
\end{eqnarray*}
where (for the case of $N=$even, $M=$odd)
\begin{eqnarray*}
{{1}\over{2}}+I_k&=&{\rm successive\;integers\;from}\;1-{1\over2}N\;\;
{\rm to}\;+{1\over2}N, \\
J_\Lambda&=&{\rm successive\;integers\;from}\;-{1\over2}(M-1)\;{\rm to}\;
+{1\over2}(M-1).
\end{eqnarray*}
We can now approach the limit $N\longrightarrow \infty$,
$M\longrightarrow \infty$, $L\longrightarrow \infty$ proportionally, obtaining
\begin{eqnarray*}
&&k=2\pi f_2-2\int_{-Q_2}^{Q_2}dk'\rho_2(k')\tan^{-1}{{(k-k')}\over{g}}
-2\int_{-B_2}^{B_2}d\Lambda\sigma_2(\Lambda)\tan^{-1}{{2(\Lambda-k)}\over{g}},
\;(A3)\\
&&2\int_{-B_2}^{B_2}d\Lambda'\sigma_2(\Lambda')
\tan^{-1}{{\Lambda-\Lambda'}\over{g}}=2\pi h_2
+2\int_{-Q_2}^{Q_2}dk\rho_2(k)\tan^{-1}{{2(\Lambda-k)}\over{g}},
\hspace{0.2in}(A4)\\
&&{{dh_2}\over{d\Lambda}}=\sigma_2,\;\;\;\;{{df_2}\over{dk}}=\rho_2,
\hspace{3.8in}(A5)\\
&&D={N\over L}=\int_{-Q_2}^{Q_2}{\rho_2(k)dk},\;\;\;
D_m={M\over L}=\int_{-B_2}^{B_2}{\sigma_2(\Lambda)d\Lambda}.
\hspace{1.7in}(A6)
\end{eqnarray*}
Or, after differentiation,
\begin{eqnarray*}
2\pi\sigma_2&=&\int_{-B_2}^{B_2}{{2g\sigma_2(\Lambda')d\Lambda'}\over
{g^2+(\Lambda-\Lambda')^2}}
-\int_{-Q_2}^{Q_2}{{4g\rho_2(k)dk}\over
{g^2+4(k-\Lambda)^2}},\hspace{1.7in}(A7)\\
2\pi\rho_2&=&1
-\int_{-B_2}^{B_2}{{4g\sigma_2(\Lambda)d\Lambda}\over
{g^2+4(\Lambda-k)^2}}+\int_{-Q_2}^{Q_2}{{2g\rho_2(k')dk'}\over
{g^2+(k-k')^2}},\hspace{1.5in}(A8)
\end{eqnarray*}
which are just eqs (54) and (55).

%\newpage
%{\Large\bf Caption}
%\vskip1cm
%\begin{description}
%\item[Fig.1] \ The curve for the function of $m$ to $e$ (eq.(38)).  Both $m$
%and $e$ are the parameters in the model (eq.(8)).
%\end{description}
\end{document}